\documentclass[
  aps,
  prl,
  reprint,
  superscriptaddress,
  nofootinbib,
  floatfix,
  nolongbibliography
]{revtex4-2}
\usepackage{framed}
\usepackage{graphicx}
\usepackage{xfrac}
\usepackage{amsmath}
\usepackage{epsfig}
\usepackage{helvet}
\usepackage{amssymb}
\usepackage{soul}
\usepackage{xcolor}
\usepackage{framed}
\usepackage[
  colorlinks=true,
  linkcolor=blue,
  citecolor=blue,
  urlcolor=blue
]{hyperref}
\usepackage{amsthm}
\usepackage{esint}
\usepackage{mathtools}

\usepackage{ragged2e}

\usepackage{tikz}
\usetikzlibrary{shapes,arrows,calc}
\usepackage{subcaption}
\usepackage[scr=boondoxupr]{mathalpha}

\begin{document}
\title{Full Eigenstate Thermalization in Quantum Field Theory \\ and Gravitational Scrambling}
\author{Ricardo Espíndola}
\affiliation{Institute for Advanced Study, Tsinghua University, Beijing 100084, China}
\author{Viktor Jahnke}
\affiliation{Instituto de Física Teórica, UNESP-Universidade Estadual Paulista
R. Dr. Bento T. Ferraz 271, Bl. II, Sao Paulo 01140-070, SP, Brazil}

\vskip 0.15cm

\begin{abstract}
\noindent
Using thermal analyticity, we derive directional large-frequency bounds
on the smooth functions entering full eigenstate thermalization in
continuum quantum field theory. The strongest directions reproduce the
scales previously identified for lattice systems, while the complete
analyticity domain reveals an asymmetric multifrequency structure.
Thermal conformal correlators saturate the second- and third-order
bounds, and the leading gravitational Regge contribution saturates the
strongest fourth-order direction.

\end{abstract}

\maketitle

\textit{\bf 1. Introduction.} Since the foundational works identifying
out-of-time-order correlators (OTOCs) as diagnostics of scrambling in
holographic systems
\cite{Shenker:2013pqa,Shenker:2013yza,Shenker:2014cwa,
Roberts:2014isa,Roberts:2014ifa}, together with the derivation of the
universal bound on chaos \cite{Maldacena:2015waa}, quantum chaos has
played an increasingly important role in AdS/CFT
\cite{Maldacena:1997re}, quantum many-body physics
\cite{Garcia-Mata:2022voo}, and quantum information
\cite{Hosur:2015ylk}. In particular, the universal properties of chaotic
systems provide powerful guiding principles for constraining theories
of quantum gravity \cite{Altland:2026tog}. Most investigations of scrambling, however, focus on four-point OTOCs
and the associated Lyapunov exponent. It is therefore natural to ask
whether quantum chaos imposes a broader hierarchy of constraints on
higher-order OTOCs. Such correlators are particularly important for
characterizing the emergence of freeness, the natural notion of
statistical independence for noncommuting observables
\cite{Jindal:2024zcg,Fava:2023pac}. In classical chaotic dynamics,
observables separated by sufficiently long times may become
statistically independent \cite{Gesteau:2023rrx,Ouseph:2023juq}. In
quantum systems, where observables do not generally commute, a natural
noncommutative analogue is freeness \cite{Camargo:2025zxr}. This
structure is characterized by the vanishing of mixed free cumulants,
or equivalently by the factorization of mixed moments according to
noncrossing partitions. Describing this hierarchy of higher-order
correlations microscopically requires control over correlations among
multiple matrix elements in the energy eigenbasis.

Such information is provided by the Eigenstate Thermalization
Hypothesis (ETH). While standard ETH
\cite{Srednicki:1994mfb,Deutsch:1991msp} determines the leading
statistical properties of individual matrix elements, it does not
completely specify the correlations among multiple matrix elements
required to describe generic higher-point functions. Full ETH
\cite{Foini:2018sdb} extends the ansatz to this complete hierarchy of
correlations, including higher-order OTOCs, and admits a natural
formulation in terms of free cumulants and free probability theory
\cite{Pappalardi:2022aaz}. So far, this framework has been developed
primarily for lattice systems. Its implications for continuum quantum
field theory, where local correlation functions are affected by
ultraviolet singularities, and for gravitational scrambling remain
comparatively unexplored.

\smallskip
\textit{In this work}, we develop full ETH in continuum quantum field
theory and holography. Using the complex-time analyticity of regulated
thermal free cumulants, we derive an infinite hierarchy of directional
bounds on the large-frequency behavior of the smooth functions entering
the full-ETH ansatz. The strongest directions reproduce the exponential
scales previously identified for lattice systems
\cite{Murthy:2019fgs,Pappalardi:2022aaz}, while the complete
analyticity domain reveals an asymmetric multifrequency structure.
Because our derivation does not rely on the finiteness of equal-time
correlators, it remains applicable to local operators in continuum quantum field theory (QFT).
We show that in two-dimensional thermal conformal field theory (CFT) two- and three-point
functions saturate the corresponding bounds, as dictated by their
conformally fixed kinematic structure. Genuinely nonuniversal behavior
first appears at fourth order, where the correlator depends on
dynamical CFT data. In this case, we find saturation of the strongest
directional bound for holographic CFTs through the leading
gravitational Regge contribution. This provides nontrivial evidence
that gravitational scrambling realizes full-ETH constraints through
fourth order and motivates the conjecture of analogous saturation at
higher orders.

\smallskip
\textit{\bf 2. Full ETH and thermal free cumulants.} The standard Eigenstate Thermalization Hypothesis \cite{Deutsch:1991msp,Srednicki:1994mfb} provides an ansatz
for the matrix elements of simple operators in the energy eigenbasis
and correctly describes equilibrium one- and two-point
functions, but it does not fully specify the correlations among several
matrix elements that enter generic higher-point functions. The full
Eigenstate Thermalization Hypothesis \cite{Foini:2018sdb} extends this framework by
postulating an ansatz for the complete hierarchy of such correlations.

When all energy indices are distinct, full ETH assumes
\begin{equation}
\overline{
O_{1,i_1i_2}O_{2,i_2i_3}\cdots O_{n,i_ni_1}
}
=
e^{-(n-1)S(E_+)}
F^{(n)}_{E_+}(\vec{\omega})\,,
\label{eq:fullETH}
\end{equation}
where
\begin{equation}
E_+=\frac{1}{n}\sum_{a=1}^{n}E_{i_a},
\qquad
\omega_a=E_{i_a}-E_{i_{a+1}},
\qquad
i_{n+1}\equiv i_1,
\end{equation}
and $\vec{\omega}$ denotes the $n-1$ independent energy differences, which satisfy $\sum_{a=1}^{n}\omega_a=0$.
Here, $S(E_+)$ is the thermodynamic entropy, while
$F^{(n)}_{E_+}(\vec{\omega})$ is a smooth function encoding the
irreducible correlations among $n$ matrix elements. Full ETH also specifies contributions involving repeated indices. Whenever an index repetition decomposes the original cyclic structure
into disconnected closed loops, the corresponding average factorizes
\begin{align}
&\overline{
O_{1,i_1i_2}\cdots O_{k-1,i_{k-1}i_1}
O_{k,i_1i_{k+1}}\cdots O_{n,i_ni_1}
}
\nonumber\\
&\qquad =
\overline{
O_{1,i_1i_2}\cdots O_{k-1,i_{k-1}i_1}
}\,
\overline{
O_{k,i_1i_{k+1}}\cdots O_{n,i_ni_1}
}.
\label{eq:factorization}
\end{align}

A particularly simple formulation of full ETH is obtained using
free probability \cite{Pappalardi:2021ahe}. Whenever full ETH holds, the smooth functions $F^{(n)}$ are
directly related to thermal free cumulants through a multifrequency
Fourier transform. Recall that free cumulants $\kappa_n$ are defined recursively through the
moment--cumulant relation
\begin{equation}
\left\langle O_1O_2\cdots O_n\right\rangle
=
\sum_{\pi\in NC(n)}
\prod_{B\in\pi}
\kappa_{|B|}
\bigl(O_{B(1)},\ldots,O_{B(|B|)}\bigr),
\label{eq:moment-cumulant}
\end{equation}
where $NC(n)$ denotes the set of noncrossing partitions of
$\{1,\ldots,n\}$. For each partition $\pi$, $B$ denotes one of its
blocks, $|B|$ is the number of elements in that block, and
$B(1)<\cdots<B(|B|)$ are its elements written in their natural order.

The factorization properties of full ETH imply that the $n$-th free
cumulant isolates the irreducible cyclic contribution with distinct
energy indices. After performing the thermal saddle, one obtains
\begin{equation}
\kappa_n^\beta
\bigl(O_1(t_1),\ldots,O_n(t_n)\bigr)
=
\int d\vec{\omega}\,
F^{(n)}_{E_\beta}(\vec{\omega})\,
e^{i\vec{\omega}\cdot\vec{t}}\,
e^{-\beta\vec{\omega}\cdot\vec{\ell}_n},
\label{eq:thermal-cumulant}
\end{equation}
where $E_\beta=\langle H\rangle_\beta$ and
$
\vec{\ell}_n
=
\left(
\frac{n-1}{n},
\frac{n-2}{n},
\ldots,
\frac{1}{n}
\right).
$
Equation \eqref{eq:thermal-cumulant} establishes a one-to-one
correspondence between the hierarchy of full-ETH smooth functions and
the hierarchy of thermal free cumulants.

\smallskip
\textit{\bf 3. Bounds from thermal analyticity.}
Lattice derivations of full-ETH bounds rely on the finiteness of
equal-time free cumulants and therefore do not apply directly to local
operators in continuum QFT. We instead regulate the cumulants by
placing the operators at equal intervals along the Euclidean thermal
circle. Defining
\begin{equation}
y_n=\frac{e^{-\beta H/n}}{Z^{1/n}},
\qquad
F_n^{\mathrm{reg}}(\vec t)
=
\operatorname{Tr}\!\left[
y_nO_1(t_1)\cdots y_nO_n(t_n)
\right],
\end{equation}
and extracting its free-cumulant component, full ETH gives
\begin{equation}
\kappa_n^{\mathrm{reg}}(\vec t)
=
\int d\vec\omega\,
F^{(n)}_{E_\beta}(\vec\omega)\,
e^{i\vec\omega\cdot\vec t}.
\label{eq:regulated-cumulant-FT}
\end{equation}

Using time-translation invariance, we set $t_n=0$ and deform the remaining contours according to $t_a\rightarrow t_a+i\eta_a$. The Euclidean separations between consecutive insertions are
\begin{equation}
a_1=\frac{\beta}{n}+\eta_1,\qquad
a_j=\frac{\beta}{n}-\eta_{j-1}+\eta_j,
\qquad
a_n=\frac{\beta}{n}-\eta_{n-1},
\end{equation}
where $j=2,\ldots,n-1$. Thermal analyticity therefore holds in the domain
\begin{equation}
\mathcal D_n
=\left\{ \vec\eta\in\mathbb R^{n-1}:a_j>0,\
j=1,\ldots,n \right\}.
\label{eq:analyticity-simplex}
\end{equation}
The inverse Fourier transform of Eq.~\eqref{eq:regulated-cumulant-FT} can be written as
\begin{equation}
F^{(n)}_{E_\beta}(\vec\omega)
=\int\frac{d^{n-1}t}{(2\pi)^{n-1}}\,
e^{-i\vec\omega\cdot\vec t}\,
\kappa_n^{\mathrm{reg}}(\vec t).
\label{eq:inverse-fourier-Fn}
\end{equation}
For any fixed $\vec\eta\in\mathcal D_n$, analyticity allows the integration contours to be shifted without crossing singularities, giving
\begin{equation}
F^{(n)}_{E_\beta}(\vec\omega)
=
\int\frac{d^{n-1}t}{(2\pi)^{n-1}}\,
e^{-i\vec\omega\cdot(\vec t+i\vec\eta)}
\kappa_n^{\mathrm{reg}}(\vec t+i\vec\eta)\,.
\label{eq:deformed-Fourier}
\end{equation}
Taking the absolute value then yields
\begin{equation}
\left|
F^{(n)}_{E_\beta}(\vec\omega)
\right|
\leq
C(\vec\eta)\,
e^{\vec\omega\cdot\vec\eta}\,,\,\,\,\,\,\,
C(\vec\eta)
\equiv
\int\frac{d^{n-1}t}{(2\pi)^{n-1}}\,
\left|
\kappa_n^{\mathrm{reg}}(\vec t+i\vec\eta)
\right|.
\label{eq:contour-bound}
\end{equation}

At this stage we make an explicit boundedness assumption: for contours in the interior of $\mathcal D_n$, $C(\vec\eta)$ is finite, and as the contour approaches a boundary corresponding to an operator collision, it grows at most algebraically in the inverse Euclidean separation. In a local QFT this behavior is naturally suggested by the operator-product expansion, while sufficient real-time clustering is assumed to control the integral away from the collision regions. In particular, $C(\vec\eta)$ must not generate an additional exponential dependence on the large-frequency scale considered below. This assumption plays a role analogous to the boundedness condition entering the derivation of the conventional chaos bound in \cite{Maldacena:2015waa}.

Writing $\vec\omega=W\vec v$, with $W\rightarrow\infty$ and $\vec v$ fixed, Eq.~\eqref{eq:contour-bound} becomes
\begin{equation}
\left|
F^{(n)}_{E_\beta}(W\vec v)
\right|
\leq
C(\vec\eta)\,
e^{W\vec v\cdot\vec\eta}.
\label{eq:directional-prebound}
\end{equation}
The optimal exponential suppression is obtained by approaching the boundary of the allowed analyticity domain in the direction that minimizes $\vec v\cdot\vec\eta$. Since the boundary singularities contribute only algebraic factors under the assumption above, one obtains
\begin{equation}
\left|
F^{(n)}_{E_\beta}(W\vec v)
\right|
\lesssim
\operatorname{poly}(W)\,
e^{-W\alpha_n(\vec v)}\,,
\,\,\,\,
\alpha_n(\vec v)
=-\inf_{\vec\eta\in\mathcal D_n}
\vec v\cdot\vec\eta.
\label{eq:QFT-directional-bound}
\end{equation}
as $W\rightarrow\infty$. Thus, the exponential large-frequency decay is determined geometrically by the support function of the thermal analyticity domain, while short-distance singularities affect only the polynomial prefactor.

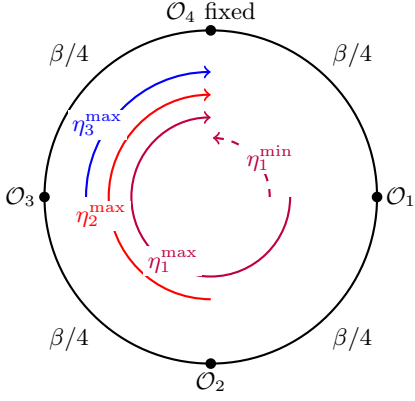
\begin{figure}[t]
\centering
\begin{tikzpicture}[scale=1.0]

\draw[thick] (0,0) circle (2.2);

\coordinate (O4) at (90:2.2);
\coordinate (O3) at (180:2.2);
\coordinate (O2) at (270:2.2);
\coordinate (O1) at (0:2.2);

\fill (O4) circle (2pt);
\fill (O3) circle (2pt);
\fill (O2) circle (2pt);
\fill (O1) circle (2pt);

\node[above] at (O4) {$\mathcal O_4$ fixed};
\node[left] at (O3) {$\mathcal O_3$};
\node[below] at (O2) {$\mathcal O_2$};
\node[right] at (O1) {$\mathcal O_1$};

\node at (135:2.65) {$\beta/4$};
\node at (225:2.65) {$\beta/4$};
\node at (315:2.65) {$\beta/4$};
\node at (45:2.65) {$\beta/4$};

\draw[->, thick, blue]
  (180:1.65) arc[start angle=180,end angle=90,radius=1.65];

\draw[->, thick, red]
  (270:1.35) arc[start angle=270,end angle=90,radius=1.35];

\draw[->, thick, purple]
  (0:1.05) arc[start angle=0,end angle=-270,radius=1.05];

\node[blue, fill=white, inner sep=1.5pt] at (-1.50,0.95) {$\eta_3^{\max}$};
\node[red, fill=white, inner sep=1.5pt] at (-1.45,-0.25) {$\eta_2^{\max}$};
\node[purple, fill=white, inner sep=1.5pt] at (-0.5,-0.85) {$\eta_1^{\max}$};


\draw[->, thick, dashed, purple]
  (0:0.78) arc[start angle=0,end angle=88,radius=0.78];

  \node[purple, fill=white, inner sep=1.2pt] at (0.78,0.52) {$\eta_1^{\min}$};

\end{tikzpicture}
\caption{ \justifying
Imaginary-time deformations of four equally spaced operator
insertions on the thermal circle. Fixing $\mathcal O_4$ by
time-translation invariance, the boundary of the analyticity domain
allows maximal positive displacements
$\eta_3^{\max}=\beta/4$,
$\eta_2^{\max}=\beta/2$, and
$\eta_1^{\max}=3\beta/4$ while preserving the cyclic ordering.
For $\mathcal O_1$, the opposite boundary is
$\eta_1^{\min}=-\beta/4$ (dashed line). Hence the allowed displacement satisfies
$-\beta/4<\eta_1<3\beta/4$, which implies
$\alpha_4(+\hat e_1)=\beta/4$ and
$\alpha_4(-\hat e_1)=3\beta/4$.
}
\label{fig:kms-four-point-shift}
\end{figure}

As an illustrative example, consider the two coordinate rays
$\vec v=\pm\hat e_1$, corresponding to
$\omega_1=\pm W$ with the remaining independent frequencies fixed.
The projection of $\mathcal D_n$ onto the $\eta_1$ axis is
\begin{equation}
-\frac{\beta}{n}
<
\eta_1
<
\frac{(n-1)\beta}{n}.
\end{equation}
For $n=4$, the corresponding imaginary-time deformations are
illustrated in Fig.~\ref{fig:kms-four-point-shift}: fixing
$\mathcal O_4$, the two boundaries are
$\eta_1^{\min}=-\beta/4$ and
$\eta_1^{\max}=3\beta/4$.
Consequently,
\begin{equation}
\alpha_n(+\hat e_1)=\frac{\beta}{n},
\qquad
\alpha_n(-\hat e_1)=\frac{(n-1)\beta}{n},
\label{eq:coordinate-rates}
\end{equation}
and hence
\begin{equation}
\left|
F^{(n)}_{E_\beta}(\vec\omega)
\right|
\lesssim
\begin{cases}
\operatorname{poly}(\omega_1)
e^{-\beta\omega_1/n},
&
\omega_1\rightarrow+\infty,
\\[2mm]
\operatorname{poly}(|\omega_1|)
e^{-(n-1)\beta|\omega_1|/n},
&
\omega_1\rightarrow-\infty.
\end{cases}
\label{eq:coordinate-bounds}
\end{equation}
For $n>2$, the two frequency directions are therefore inequivalent.
The stronger negative-frequency scale $(n-1)\beta/n$ reproduces the
scale emphasized in previous lattice analyses \cite{Pappalardi:2022aaz}, while the complete
continuum-QFT result also resolves the weaker opposite direction. More generally, for $\omega_k$ with $k=1,\ldots,n-1$, one finds
\begin{equation}
\left|
F^{(n)}_{E_\beta}(\vec \omega)
\right|
\lesssim
\begin{cases}
\operatorname{poly}(\omega_k)
e^{-k\beta\omega_k/n},
& \omega_k\to+\infty,\\[2mm]
\operatorname{poly}(|\omega_k|)
e^{-(n-k)\beta|\omega_k|/n},
& \omega_k\to-\infty,
\end{cases}
\label{eq:QFT-general-coordinate-bound}
\end{equation}
with all other independent frequencies held fixed.

\smallskip
\textit{ \bf 4. Saturation in CFT and gravity.}  In a two-dimensional CFT, thermal two- and three-point functions on the
cylinder are fixed by conformal symmetry up to operator normalizations,
scaling dimensions, and OPE coefficients. They therefore provide
model-independent tests of the second- and third-order bounds. The
four-point function is the first correlator containing nontrivial
dynamical information through its dependence on conformal cross-ratios.
We first demonstrate saturation for the universal two- and three-point
structures and then turn to the leading Regge contribution in
holographic CFTs.

\textit{Kinematic saturation in thermal CFT$_2$.}
We first consider spatially coincident scalar primary operators with
vanishing thermal one-point functions, for which the second and third
free cumulants coincide with the corresponding correlation functions.
For two equally spaced insertions,
\begin{equation}
\kappa^{\mathrm{reg}}_2(t)
=
C_O
\left(\frac{\pi}{\beta}\right)^{2\Delta}
\frac{1}{\cosh^{2\Delta}(\pi t/\beta)}.
\label{eq:CFT2two}
\end{equation}
Its exact Fourier transform is proportional to
$\left|\Gamma(\Delta+i\beta\omega/2\pi)\right|^2$, and therefore
\begin{equation}
\widehat{\kappa}_2(\pm W)
\sim
W^{2\Delta-1}e^{-\beta W/2}.
\label{eq:CFT2twoAsymptotic}
\end{equation}
The two frequency directions coincide, in agreement with
$\alpha_2(\pm\hat e_1)=\beta/2$.

\begin{figure*}[t]
    \centering
    \includegraphics[width=\textwidth]{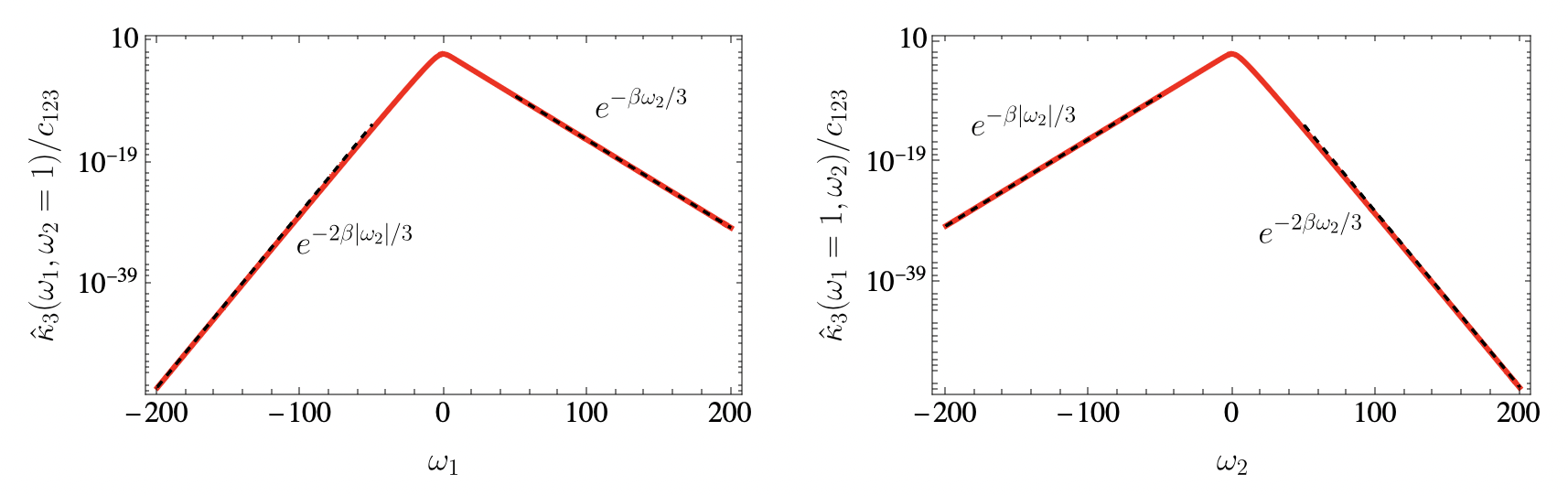}
    \caption{\protect\justifying
    Large-frequency behavior of the Fourier transform of the third regulated
    free cumulant $\widehat{\kappa}_3(\omega_1,\omega_2)$ for $\beta=1$ and
    $\Delta_1=\Delta_2=\Delta_3=1$. Left: $\omega_2=1$ with
    $\omega_1\rightarrow\pm\infty$. Right: $\omega_1=1$ with
    $\omega_2\rightarrow\pm\infty$. Black dashed lines show the directional
    exponential decay rates predicted by Eq.~\eqref{eq:QFT-general-coordinate-bound}.}
    \label{fig:kappa3}
\end{figure*}

For three scalar primaries, let
$\delta_{ab}=\Delta_a+\Delta_b-\Delta_c$, with $c\neq a,b$, and define the shifted kernel
\begin{equation}
\mathcal K_{\delta,a}(t)
\equiv
\left[i
\sinh\!\left(
\frac{\pi}{\beta}(t-ia)
\right)
\right]^{-\delta}.
\label{eq:thermal-kernel}
\end{equation}
Setting $t_3=0$, the equally spaced regulated correlator becomes
\begin{align}
\kappa^{\mathrm{reg}}_3(t_1,t_2)
={}&
c_{123}
\left(\frac{\pi}{\beta}\right)^{
\Delta_1+\Delta_2+\Delta_3}
\mathcal K_{\delta_{12},\beta/3}(t_1-t_2)
\nonumber\\
&\times
\mathcal K_{\delta_{23},\beta/3}(t_2)
\mathcal K_{\delta_{13},2\beta/3}(t_1).
\label{eq:CFT2three}
\end{align}
Its double Fourier transform can be written as
\begin{align} \label{eq:convolution}
\widehat{\kappa}_3(\omega_1,\omega_2)={}&\frac{c_{123}}{2\pi}  \left(\frac{\pi}{\beta}\right)^{\Delta_1+\Delta_2+\Delta_3} \int_{-\infty}^{\infty}d\omega \, \hat{K}_{\delta_{12},\beta/3}(\omega)\nonumber \\
&\hat{K}_{\delta_{23},\beta/3}(\omega_2+\omega)\,\hat{K}_{\delta_{13},2\beta/3}(\omega_1-\omega)\,,
\end{align}
where
\begin{equation} 
\widehat{K}_{\delta,a}(\omega)
=e^{-\left(\frac{\beta}{2}-a\right)\omega}
\frac{\beta}{\pi}\,
\frac{2^{\delta-1}}{\Gamma(\delta)}
\Gamma\left(
\frac{\delta}{2}+\frac{i\beta\omega}{2\pi}
\right)
\Gamma\left(
\frac{\delta}{2}-\frac{i\beta\omega}{2\pi}
\right)\,.
\label{eq:CFT-kernel-FT-def}
\end{equation}
The convolution formula \eqref{eq:convolution} allows us to study numerically the large-frequency behavior of $\widehat{\kappa}_3(\omega_1,\omega_2)$ and compare it with the directional decay rates predicted by the full-ETH bounds. Figure~\ref{fig:kappa3} illustrates this behavior for
$\Delta_1=\Delta_2=\Delta_3=1$, with $\omega_2$ fixed and
$|\omega_1|$ large in the left panel, and with $\omega_1$ fixed and
$|\omega_2|$ large in the right panel. In both cases, the full-ETH
bound \eqref{eq:QFT-general-coordinate-bound} is saturated. The same
directional exponential rates can be shown to hold for generic
nondegenerate scaling dimensions.
We next show that the strongest fourth-order direction is realized by gravitational scrambling.

\textit{Saturation from gravitational scrambling.}
Unlike two- and three-point functions, the CFT four-point function contains
genuine dynamical information. We now show that, at leading semiclassical
order in the Regge regime, the gravitational contribution to the fourth
free cumulant saturates the strongest directional bound. For simplicity, we first consider the AdS$_2$/CFT$_1$ eikonal result \cite{Maldacena:2016upp}, where there are no transverse directions and the shock-wave profile can be absorbed into the effective eikonal coupling $g$. The role of transverse dynamics in higher-dimensional backgrounds is discussed in the companion paper \cite{companion}.

Consider two centered scalar primaries $V$ and $W$, of dimensions
$\Delta_V$ and $\Delta_W$, and the regulated correlator
\begin{equation}
F_4(\vec t)
=
\operatorname{Tr}
\left[
yV(t_1)yW(t_2)yV(t_3)yW(t_4)
\right],
\,\,\,\,
y=\frac{e^{-\beta H/4}}{Z^{1/4}}\,.
\end{equation}
For centered operators, the fourth free cumulant is
\begin{equation}
\kappa_4(1,2,3,4)
=
F(1,2,3,4)
-F(1,2)F(3,4)
-F(1,4)F(2,3)\,.
\end{equation}
The ordinary connected correlator, by contrast, is obtained by subtracting
all three pairwise factorizations. It follows that
\begin{equation}
\kappa_4(1,2,3,4)
=
F_{\rm conn}(1,2,3,4)
+
F(1,3)F(2,4)\,.
\end{equation}
Thus, the fourth free cumulant contains the connected contribution together
with the crossing factorized term. This is precisely the combination
naturally captured by the holographic eikonal channel: in the absence of
gravitational scattering it reduces to $F(1,3)F(2,4)$, while the nontrivial
eikonal phase generates the connected gravitational contribution.

In the Regge regime, the states created by the two pairs of boundary
insertions can be decomposed into null-momentum eigenstates on the future
and past horizons. The corresponding gravitational scattering amplitude
is the eikonal phase $e^{i\delta}$, with $\delta \propto g \,pq$. After continuation to the equally spaced thermal
contour, the fourth free cumulant can therefore be written as
\begin{equation}
 \kappa_4(\vec t)
 =
 \int_0^\infty dp\,dq\,
 \Phi_V(p;t_1,t_3)\,
 \Phi_W(q;t_2,t_4)\,
 e^{- g p q},
 \label{eq:eikonal-general}
\end{equation}
where $\Phi_V$ and $\Phi_W$ are products of horizon wavefunctions
obtained from the corresponding Rindler--AdS bulk-to-boundary
propagators. Evaluating these wavefunctions for scalar primaries gives
\begin{align}
\kappa_{4}(\vec t)
={}&
\frac{\mathcal N\,P_V(t_{13})P_W(t_{24})}
{\Gamma(2\Delta_V)\Gamma(2\Delta_W)}
\int_0^\infty dp\,dq\,
p^{2\Delta_V-1}q^{2\Delta_W-1}e^{-p-q}
\nonumber\\
&\times
e^{-g pq\,\mathcal R(\vec t)},
\label{eq:resolved-regge}
\end{align}
where $t_{ij}=t_i-t_j$,
$P_X(t)=[2\cosh(\kappa t/2)]^{-2\Delta_X}$, and
\begin{equation}
\mathcal R(\vec t)
=
\frac{4}{
(e^{\kappa t_1}+e^{\kappa t_3})
(e^{-\kappa t_2}+e^{-\kappa t_4})
},
\qquad
\kappa=\frac{2\pi}{\beta}.
\label{eq:resolved-regge-ratio}
\end{equation}
Here $g$ is the effective eikonal coupling. Equation~\eqref{eq:resolved-regge} follows from the
standard horizon-wavefunction representation of the gravitational
eikonal amplitude; its detailed derivation will be presented in a
companion paper \cite{companion}.

Using time-translation invariance, we set $t_4=0$ and define
\begin{equation}
\widehat \kappa_{4}(\vec\omega)
=
\int dt_1dt_2dt_3\,
e^{-i(\omega_1t_1+\omega_2t_2+\omega_3t_3)}
\kappa_{4}(t_1,t_2,t_3,0),
\label{eq:regge-multift}
\end{equation}
with $\omega_4=-\omega_1-\omega_2-\omega_3$. The change of variables
\begin{equation}
t_u=\frac{t_3-t_1}{2},
\qquad
t_v=\frac{t_2}{2},
\qquad
t_s=\frac{t_2-t_1-t_3}{2}
\label{eq:regge-variables}
\end{equation}
separates the relative distances within the two operator pairs from
their relative Regge boost. In particular,
\begin{equation}
\mathcal R
=
\frac{e^{\kappa t_s}}
{\cosh(\kappa t_u)\cosh(\kappa t_v)}~.
\end{equation}
A subtlety of the multitime transform is that the real times in
Eq.~\eqref{eq:regge-multift} are integrated independently over the full real line and
therefore include regions that are not out of time order in the
literal real-time sense, even though the imaginary parts of the times keep track of the ordering of the operators around the thermal circle. More importantly, although the Fourier transform formally samples the full
real-time domain, its large-frequency Regge contribution is controlled
by increasingly large relative boosts, precisely where the eikonal
description applies.

Performing first the integral over $t_s$, then the integrals over $p$ and $q$, and finally the integrals over $t_u$ and $t_v$, we obtain
\begin{align}
\widehat{\kappa}_{4}(\vec\omega)
={}&
\frac{\mathcal N}{\kappa^3}
(4g)^{-i\nu}
\Gamma(i\nu)
\Gamma\!\left(
\Delta_V-\frac{i\omega_1}{\kappa}
\right)
\Gamma\!\left(
\Delta_V-\frac{i\omega_3}{\kappa}
\right)
\nonumber\\
&\times
\Gamma\!\left(
\Delta_W+\frac{i\omega_2}{\kappa}
\right)
\Gamma\!\left(
\Delta_W+\frac{i\omega_4}{\kappa}
\right),
\label{eq:regge-full-transform}
\end{align}
where
\begin{equation}
\nu=\frac{\omega_1+\omega_3}{\kappa},
\qquad
\omega_4=-\omega_1-\omega_2-\omega_3.
\end{equation}

To study the large-frequency behavior of the expression, it is useful to employ the Stirling asymptotic formula, $|\Gamma(a+i\nu)|
\sim
\sqrt{2\pi}\,
|\nu|^{a-\frac12}
e^{-\frac{\pi}{2}|\nu|},
$ as $
|\nu|\rightarrow\infty.$ Applying this formula to the frequency-dependent Gamma functions in the expression above, while keeping the remaining independent frequencies fixed, we find
\begin{equation}
\left|\widehat{\kappa}_4(\vec{\omega})\right|
\sim
\operatorname{poly}(W)
\begin{cases}
e^{-3\beta W/4}\,, & \omega_1=\pm W\,,\\[2pt]
e^{-\beta W/2}\,, & \omega_2=\pm W\,,\\[2pt]
e^{-3\beta W/4}\,, & \omega_3=\pm W\,,
\end{cases}
\qquad W\rightarrow\infty\,.
\label{eq:kappa4-coordinate-asymptotics}
\end{equation}
Comparing with Eq.~\eqref{eq:QFT-general-coordinate-bound}, the
strongest coordinate bounds are therefore saturated for
$\omega_1\rightarrow-\infty$, $\omega_2\rightarrow\pm\infty$, and
$\omega_3\rightarrow+\infty$, while the opposite $\omega_1$ and
$\omega_3$ directions decay faster than required. This behavior follows from the symmetry of the holographic result,
\begin{equation}
    \widehat{\kappa}_4(-\vec{\omega})
    =
    \widehat{\kappa}_4(\vec{\omega})^*
    \,\,\Longrightarrow \,\,
    \left|\widehat{\kappa}_4(-\vec{\omega})\right|
    =
    \left|\widehat{\kappa}_4(\vec{\omega})\right|\,,
\end{equation}
which implies that once the stronger bound is saturated in one frequency
direction, the opposite direction must exhibit the same exponential
decay rate, and therefore decays faster than required by the weaker
bound. The same frequency dependence applies to fourth-order free cumulants in Rindler–AdS$_3$. We discuss the details of higher-dimensional cases in a companion paper.

\smallskip
\textit{\bf 5. Discussion.}
Assuming full ETH, regulated thermal free cumulants are multifrequency Fourier transforms of the corresponding ETH smooth functions. Their complex-time analyticity domains, together with the at-most power-law singularities associated with coincident operator insertions, constrain the exponential decay of these functions at large frequencies. Along the strongest directions, the resulting scales reproduce those previously identified for lattice systems \cite{Murthy:2019fgs,Pappalardi:2022aaz}, while the full analyticity domain reveals a more general directional and asymmetric structure. In thermal CFT$_2$, the universal two- and three-point functions saturate the corresponding bounds, whereas genuinely dynamical information first enters at fourth order through the dependence on conformal cross-ratios.

The holographic fourth-order result is therefore particularly nontrivial. In the eikonal regime, the gravitational contribution saturates the strongest directional bound, providing evidence that gravitational scrambling realizes the full-ETH constraints beyond the kinematically fixed two- and three-point sectors. This bound is also closely related to the conventional chaos bound: as shown in Ref.~\cite{Murthy:2019fgs}, the corresponding large-frequency constraint can be converted, for an appropriate OTOC form, into the Lyapunov bound $\lambda_L\leq2\pi/\beta$. The present hierarchy may thus be viewed as a multifrequency extension of this frequency-space manifestation of the chaos bound.

Further support for a connection between gravity and full ETH comes from the gravitational scrambling algebra of Ref.~\cite{Penington:2025hrc}, which interpolates between a tensor-product structure at early times and a free-product structure at late times. The departure from the tensor-product regime is governed by maximal Lyapunov growth, while the late-time algebra implies freeness between sufficiently separated early- and late-time observables. Full ETH likewise leads to late-time freeness in a time-averaged sense \cite{Fava:2023pac}, with fluctuations suppressed in the thermodynamic limit. Although algebraic freeness alone does not establish full ETH or imply saturation of the frequency-space bounds, this parallel structure suggests that gravitational dynamics may realize the broader hierarchy encoded by full ETH. It would therefore be interesting to test the directional bounds at higher orders, beginning with six-point OTOCs such as those studied in Ref.~\cite{Haehl:2021tft}.

More broadly, our construction provides a framework for studying full ETH directly in continuum quantum field theory. Tests of the directional bounds \eqref{eq:QFT-directional-bound} and their possible saturation could provide indirect probes of higher-order ETH correlations in systems where conventional spectral diagnostics are subtle. They may also provide constraints on effective hydrodynamic descriptions of quantum chaos.

\begin{acknowledgements}
\noindent
\textit{\bf Acknowledgements.---} We thank Bartek Czech, Jan de Boer, Ben Freivogel, Mark Mezei, Miguel Tierz, and Huajia Wang for insightful discussions. R.~Esp\'indola thanks Roberto Emparan for hospitality and useful discussions at the Institute of Cosmos Sciences of the University of Barcelona (ICCUB), where part of this work was carried out. He also thanks the organizers of “Holo-Asia 2026'' (Jeju) and of the “Amsterdam String Summer Workshop 2026''. R.~Esp\'indola is supported by the Shuimu Tsinghua Scholar Program. V.~Jahnke was supported by the Conselho Nacional de Desenvolvimento Científico e Tecnológico (CNPq), Brazil, under grant Processo 446326/2024-0 (Bolsa Conhecimento Brasil – BCB-1).
\end{acknowledgements}

\bibliography{refprl}

\end{document}